\documentclass{emulateapj}
\usepackage{mathptmx}
\shorttitle{OBSERVATIONS OF SOLAR DOPPLER SHIFT OSCILLATIONS}
\shortauthors{MARISKA ET AL.}

\slugcomment{Accepted May 23, 2008}

\begin{document}

\title{OBSERVATIONS OF DOPPLER SHIFT OSCILLATIONS WITH THE EUV
  IMAGING SPECTROMETER ON \textit{HINODE}}

\author{John T. Mariska\altaffilmark{1}, 
  Harry P. Warren\altaffilmark{1},
  David R. Williams\altaffilmark{2},
  and Tetsuya Watanabe\altaffilmark{3}}
\altaffiltext{1}{Space Science Division, Code 7673, Naval
  Research Laboratory, Washington, DC 20375}
\altaffiltext{2}{Mullard Space Science Laboratory, University
  College London, Holmbury St.\ Mary, Dorking, Surrey, RH5 6NT,
  UK}
\altaffiltext{3}{National Astronomical Observatory of Japan,
  2-21-1 Osawa, Mitaka-shi, Tokyo 181-8588, Japan}
\email{mariska@nrl.navy.mil}

\begin{abstract}
Damped Doppler shift oscillations have been observed in emission
lines from ions formed at flare temperatures with the Solar
Ultraviolet Measurements of Emitted Radiation spectrometer on the
\textit{Solar and Heliospheric Observatory} and with the Bragg
Crystal Spectrometer on \textit{Yohkoh}. This Letter reports the
detection of low-amplitude damped oscillations in coronal
emission lines formed at much lower temperatures observed with
the EUV Imaging Spectrometer on the \textit{Hinode} satellite.
The oscillations have an amplitude of about 2 km s$^{-1}$, and a
period of around 35~min. The decay times show some evidence for a
temperature dependence with the lowest temperature of formation
emission line (\ion{Fe}{12} 195.12~\AA) exhibiting a decay time
of about 43~min, while the highest temperature of formation
emission line (\ion{Fe}{15} 284.16~\AA) shows no evidence for
decay over more than two periods of the oscillation. The data
appear to be consistent with slow magnetoacoustic standing waves,
but may be inconsistent with conductive damping.
\end{abstract}

\keywords{Sun: corona --- Sun: oscillations --- Sun: UV
  radiation}

\section{INTRODUCTION}

In recent years, oscillatory phenomena have been detected in the
solar corona using spacecraft imaging instruments
\citep[e.g.,][]{Aschwanden1999,Schrijver2002} and spectroscopic
instruments \citep[e.g.,][]{Wang2002,Kliem2002,Mariska2006}.
Detection and characterization of coronal oscillatory phenomena
with spectroscopic instruments that simultaneously capture many
emission lines is particularly important, since it offers the
possibility of refining our understanding of the
temperature-dependent behavior of the oscillations. It may also
be possible to make simultaneous observations in
density-sensitive pairs of emission lines, further constraining
the characteristics of the oscillations.

The EUV Imaging Spectrometer (EIS) on \textit{Hinode} produces
stigmatic spectra in two 40~\AA\ wavelength bands centered at 195
and 270~\AA. EIS can image the Sun using 1\arcsec\ and
2\arcsec\ slits to produce line profiles and 40\arcsec\ and
266\arcsec\ slots to produce monochromatic images. When the slits
are used, the EIS mirror can be moved to construct
spectroheliograms by rastering an area of interest. This can
provide a snapshot of the thermal structure and dynamics of a
region of interest, but with typical exposure times ranging from
a few seconds to a minute or more, it is difficult to obtain high
time cadence dynamical data in this manner. When a higher cadence
is important, EIS can operate in a sit-and-stare mode in which
the slit covers a fixed location on the Sun and takes repeated
exposures. A more global context can then be provided by
preceding and/or following the sit-and-stare data with EIS
spectroheliograms. Further context is often available from
observations taken simultaneously with the \textit{Hinode} X-Ray
Telescope (XRT). An overall description of EIS is available in
\citet{Culhane2007}, XRT is described in \citet{Golub2007}, and
the \textit{Hinode} mission is described by \citet{Kosugi2007}.

In this Letter we show that damped oscillations are present in
some spectral lines observed in sit-and-stare observations taken
with EIS. While the earlier spectroscopic observations detected
oscillations in plasma at flare temperatures
\citep[e.g.,][]{Wang2003,Mariska2006}, the data presented here
show the presence of oscillations at typical active region
temperatures. The oscillations appear to be standing slow mode
magnetoacoustic waves, but the temperature dependence of the
decay times may be inconsistent with conductive damping.

\section{EIS OBSERVATIONS}

The observations discussed in this paper were centered at the
southwest limb (S09W84) on 2007 January 14. A sit-and-stare
observation began at 12:30:12~UT and consisted of 120 exposures
with the 1\arcsec\ slit, each with an exposure time of 60~s. Each
exposure covered nine data windows on the EIS detectors. This
Letter only presents the results from lines in six of those
windows. Data from the other detector windows were not suitable
for a Doppler shift analysis, either because of weak signal or
difficult to fit line profiles. Each window was 32 spectral
pixels wide and covered a height of 512\arcsec\ in the N/S
direction. EIS detector pixels are 22.3~m\AA\ wide in the
spectral direction and 1\arcsec\ wide in the spatial direction.
The measured FWHM for the instrument at 185~\AA\ is
47~m\AA\ \citep{Culhane2007}.

The EIS sit-and-stare observation was followed by a
spectroheliogram taken with the 1\arcsec\ slit and covering a
$240\arcsec \times 240\arcsec$ region. This data set contained 12
spectral windows, including all of those covered in the
sit-and-stare observation. The exposure times, however, were only
5~s, resulting in a relatively weak signal in some of the lines
over much of the spectroheliogram.

\begin{figure*}
\plotone{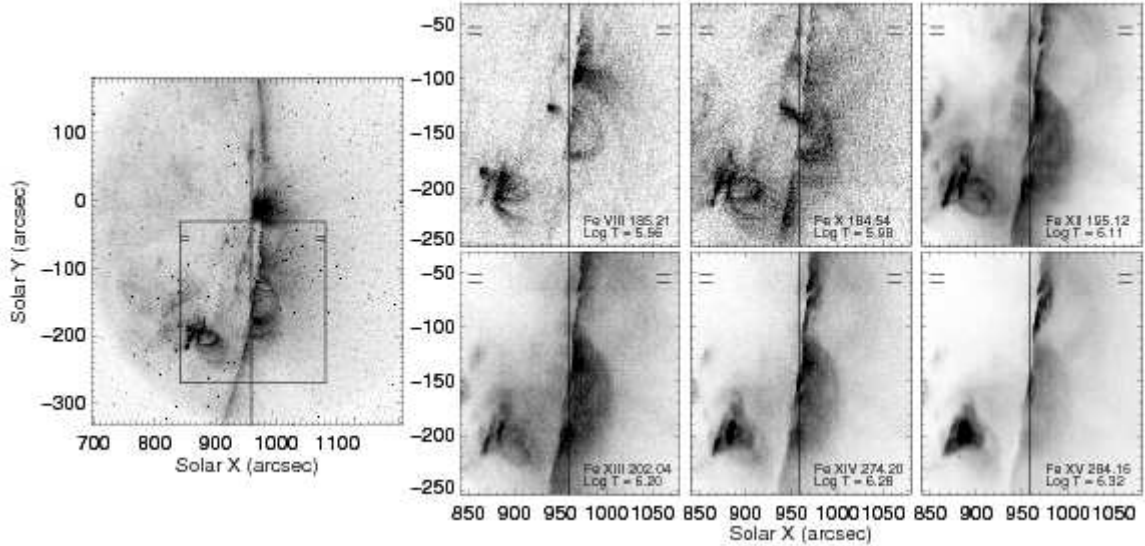}
\caption{A \textit{TRACE} 171~\AA\ image taken at 11:19:41 UT
  showing the location of the EIS sit-and-stare observation taken
  from 12:30:12 to 14:32:58 UT on 2007 January 14 and the EIS
  spectroheliograms obtained from 14:50:12 to 15:16:22 UT, which
  are shown to the right of the \textit{TRACE} image. Short
  horizontal lines on the raster outline and on each image mark
  the $y$-range of the rows in the EIS spectroheliogram that show
  Doppler-shift oscillations.}
\label{fig:context}
\end{figure*}

While the XRT was making exposures during the time of the EIS
sit-and-stare observation, the exposure times were very short,
resulting in images with significant detail only for the
strongest features. The \textit{Transition Region and Coronal
  Explorer} (\textit{TRACE}) was also observing at this location
and provided better context images. The left side of
Figure~\ref{fig:context} shows a \textit{TRACE} 171~\AA\ image of
the area around the sit-and-stare data set with the locations of
the EIS slit for the sit-and-stare observation and the following
EIS raster indicated. Also marked is the $y$-range of the
locations in the EIS raster that exhibit the oscillations
discussed in this Letter.

All the EIS data were processed to remove detector bias and dark
current, hot pixels and cosmic rays, and calibrated using the the
prelaunch absolute calibration, resulting in intensities in ergs
cm$^{-2}$ s$^{-1}$ sr$^{-1}$ \AA$^{-1}$. The EIS slit tilt and
orbital variation in the line centroids were also removed from
the data. The emission lines in each spectral window were then
fitted with Gaussian profiles, providing the total intensity,
location of the line center, and the width. The data in the long
wavelength detector were also shifted by 17 pixels in the
$y$-direction to correct for the offset between the two
detectors.

The right panels in Figure~\ref{fig:context} show the appearance
of the area covered by the EIS spectroheliogram in the lines
contained in the sit-and-star observation along with the location
of the EIS slit for the sit-and-stare observation. Beneath the
identification of each emission line, we also list the log of the
temperature of formation of the line.

Since the spacecraft was in a fixed pointing mode during both the
sit-and-stare observation and the following spectroheliogram, the
sit-and-stare slit location on the spectroheliogram is well
determined. Placing the slit on the \textit{TRACE} image is more
challenging. That placement has been accomplished using the
feature centered at an $(x,y)$ location of about $(880, -200)$.
The placement is only accurate to a few arcseconds.

As we mentioned above, because short exposure times were used,
the XRT images only showed the brightest features. Examination of
a movie of the images, however, does show flare-like brightenings
taking place in the small loop-like features at ($x$, $y$)
locations of roughly ($850$, $-150$) and ($860$, $-130$) visible
in the \ion{Fe}{12} image and those formed at higher
temperatures. One of these events, which took place at roughly
13:00~UT, resulted in a brightening in the the feature used to
co-align with the \textit{TRACE} image and appeared to show a
very weakly emitting loop that connected to the area that
exhibited oscillations. Thus, we believe that the observed
oscillations are the result of impulsive heating events and that
the sit-and-stare data are near one footpoint of the impulsively
heated loops. Note that the event at 13:00~UT can not be the one
that generated the observed oscillations, since they are already
present in the EIS data by that time.

\section{RESULTS}

\begin{figure}
\plotone{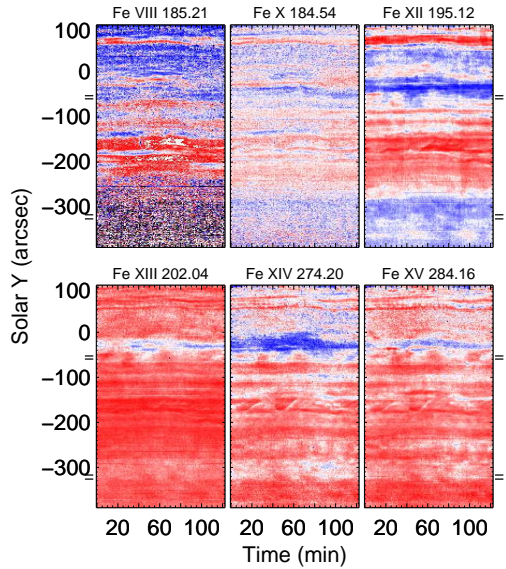}
\caption{EIS sit-and-stare Doppler-shift data in six different
  emission lines obtained beginning at 12:30:12 UT on 2007
  January 14. The horizontal axis shows the start times of the
  exposures relative to the beginning of the data set. The
  Doppler shift in each window is adjusted so that the zero value
  is the average over the window. The maximum and minimum values
  plotted in each case are $+20$ and $-20$ km s$^{-1}$,
  respectively. The short horizontal lines mark the ranges of
  rows showing Doppler-shift oscillations.}
\label{fig:eis-sns-v}
\end{figure}

Figure~\ref{fig:eis-sns-v} shows a color representation of the
measured centroid positions for each spectral line as a function
of time. For many $y$-positions, the Doppler shift appears to be
relatively constant as a function of time. There are, however, a
number of positions that exhibit time-dependent behavior. Perhaps
the most striking is the marked region near a $y$-position of
$-60\arcsec$, which in the three higher temperature lines shows a
pattern of alternating redshifted and blueshifted emission in the
Doppler-shift data and some evidence for a gradual change in the
intensity data. A second position range showing similar behavior
is indicated near the bottom of the figure.

To examine further the signal in this portion of the
Doppler-shift data, we have averaged the measured total line
intensities and centroid shifts in each detector window over the
seven rows centered on a $y$-position of $-56\arcsec$.

\begin{figure*}
\plotone{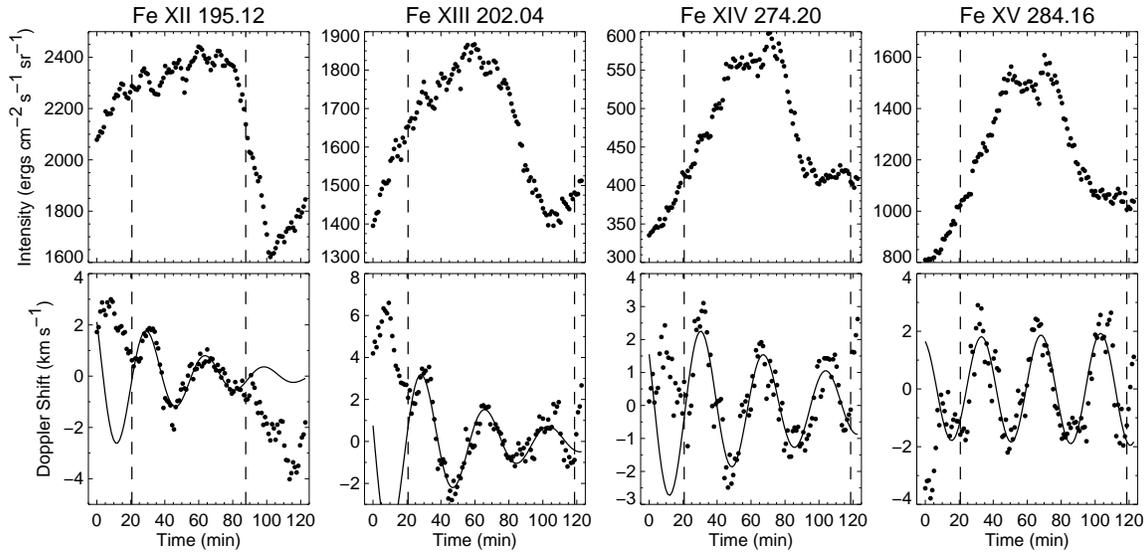}
\caption{Averaged intensity and detrended Doppler-shift data for
  the emission lines that exhibited Doppler-shift oscillations.
  The curves on the lower plots show the best-fit to a decaying
  sine wave.}
\label{fig:eis-row-osc}
\end{figure*}

The four higher temperature lines show clear evidence in the
Doppler shift averages for and oscillatory pattern, which we fit
with a combination of a polynomial background and a damped sine
wave of the form
\begin{equation}
v(t) = A_0 \sin(\omega t + \phi) \exp(-\lambda t) + B(t),
\label{eqn:sine}
\end{equation}
where
\begin{equation}
B(t) = b_0 + b_1 t + b_2 t^2 + b_3 t^3 + \dots
\end{equation}
The top and bottom panels of Figure~\ref{fig:eis-row-osc} show
the behavior of the averaged intensities and the detrended
averaged Doppler shifts in these lines, respectively. Also shown
on each of the bottom panels is the best-fit damped sine wave.
Only the data in the interval between the vertical dashed lines
was used for the fitting. Both a linear and a quadratic
background yielded similar fitting parameters for the damped sine
wave, with the quadratic fit being slightly better in some cases.
The reduced $\chi^2$ for the fits are generally around 2.

The averaged line intensities for the two cooler lines,
especially the \ion{Fe}{12} 195.12~\AA\ emission line, show some
evidence for periodic intensity fluctuations, particularly over
the first portion of the interval used for fitting the Doppler
shift data. We have, however, been unable to fit a decaying sine
wave to the data using equation (\ref{eqn:sine}).

Examination of the averaged Doppler shift data in the \ion{Fe}{8}
185.21~\AA\ emission line shows no evidence for oscillatory
behavior. The data for the \ion{Fe}{10} 184.54~\AA\ emission line
shows an indication of a small disturbance in the line centroid
at the time the oscillations are observed in the higher
temperature lines. Thus the oscillation does not appear to be
present for lines with temperatures of formation of less than
about 1 MK.

Table~\ref{table:results} summarizes the parameters of the fits
to the Doppler shift data. For all the lines in the Table, the
amplitudes, periods, and phases are roughly similar, suggesting
that we are seeing the response of the solar plasma to the same
disturbance. Note, however, that there appears to be a tendency
for the decay time to increase with increasing temperature of
formation for the line. The data for the \ion{Fe}{15} 284.16~\AA\
line are consistent with no detectable decay over the time
interval measured.

\begin{table}
\caption{Fitting Results\label{table:results}}
\begin{center}
\begin{tabular}{lllll}
\tableline
 & \multicolumn{1}{c}{$A_0$} & \multicolumn{1}{c}{$\omega$} &
\multicolumn{1}{c}{$\phi$} & \multicolumn{1}{c}{$\lambda$} \\
Line & \multicolumn{1}{c}{(km s$^{-1}$)} &
\multicolumn{1}{c}{(rad min$^{-1}$)} &
\multicolumn{1}{c}{(rad)} & \multicolumn{1}{c}{(hr$^{-1}$)} \\
\tableline
Fe~\textsc{xii} & $2.1 \pm 0.2$ & $0.181 \pm 0.003$ & $-0.1 \pm 0.1$ & 
\phs $1.4 \pm 0.2$ \\
Fe~\textsc{xiii} & $3.7 \pm 0.4$ & $0.168 \pm 0.003$ & $+0.2 \pm 0.1$ & 
\phs $1.2 \pm 0.2$ \\
Fe~\textsc{xiv} & $2.5 \pm 0.3$ &  $0.171 \pm 0.002$ & $-0.1 \pm 0.1$ & 
\phs $0.62 \pm 0.15$ \\
Fe~\textsc{xv} & $1.8 \pm 0.2$ & $0.179 \pm 0.002$ & $-0.7 \pm 0.1$ & 
$ -0.05 \pm 0.13$ \\
\tableline
\end{tabular}
\end{center}
\end{table}

\section{DISCUSSION AND CONCLUSIONS}

Defining the maximum amplitude displacement for the oscillations
\citep{Wang2003} as
\begin{equation}
A = A_0/(\omega^2 + \lambda^2)^{1/2} ,
\end{equation}
we calculate a value for $A$ in the \ion{Fe}{14} 274.20~\AA\ line
of about 860~km. Thus, if the oscillations are transverse to the
line of sight, the range of motion would only be on the order of
1\arcsec, making them very difficult to detect in images.

The Doppler shift data shown in Figure~\ref{fig:eis-row-osc}
looks similar to the Doppler shift oscillations observed with the
Solar Ultraviolet Measurements of Emitted Radiation instrument
(SUMER) on the \textit{Solar and Heliospheric Observatory}
(\textit{SOHO}) \citep[e.g.,][]{Wang2003}. Those oscillations
were primarily seen in emission lines from \ion{Fe}{19} and
\ion{Fe}{21} and appeared to be the response to flare-like
heating in active region loops. The amplitudes were much larger
than those reported here---averaging about 98 km~s$^{-1}$. The
average period reported was 17.6~min, but ranged up to 31.1~min.

As we discussed in \S2, we believe that the observed oscillation
is the result of a heating event in a nearby structure that
appears to be magnetically connected to the area where the
oscillations are observed. Assuming we are observing a loop, that
the oscillations are near one footpoint and that the other
footpoint is at an ($x$, $y$) position of approximately ($860$,
$-130$), we compute a distance between the two footpoints of
roughly 218~Mm. A semicircular loop with this diameter would then
have a length of approximately 342~Mm. For a sound speed of 200
km s$^{-1}$, the sound travel time between the footpoints is
about 29~min. This is within a factor of two of the value
expected for a standing magnetoacoustic wave oscillating in the
fundamental mode. Since the locations of the footpoints are
highly uncertain, especially since they are very close to the
limb, and the loop geometry is difficult to determine, we believe
this this level of agreement is good and suggests that we are
observing a damped standing mode slow magnetoacoustic wave.

A strong indicator that the oscillations are standing mode would
be the presence of a $1/4$ period phase shift between the
intensity and the Doppler shift \citep{Sakurai2002}. Both the
\ion{Fe}{12} and \ion{Fe}{13} line intensities show local peaks
at about 10 and 20 min into the interval used for fitting the
Doppler shift data, suggesting a period for any intensity
oscillation of about 10 min. The \ion{Fe}{14} and \ion{Fe}{15}
emission lines show local peaks at about 30 and 50 min into the
fitting interval, suggesting a period for any intensity
oscillation of of about 20 min. In both cases, these periods are
shorter than those listed in Table~\ref{table:results}. For a
sound wave, $v = c_s \delta \rho/\rho$, where $c_s$ is the sound
speed and $\rho$ is the density. Thus, for the formation
temperatures of the lines showing Doppler shift oscillations,
$\delta \rho/\rho \approx 0.01$, a very small density
fluctuation, which would not be detectable at the signal-to-noise
levels in the intensity data. Thus, we believe that the most that
can be said about the observed intensities is that they are not
inconsistent with standing mode slow magnetoacoustic waves.

The ratio of the decay time to the wave period can provide some
additional information on the damping mechanism. For the SUMER
oscillations the average value for that ratio is 0.85
\citep{Wang2003}. For the three cooler lines in
Table~\ref{table:results} the values are 1.2, 1.3, and 2.6,
respectively, with the value for the \ion{Fe}{15} line being
greater than 2.8. The result that the ratio for the \ion{Fe}{15}
line is potentially much larger than that for the other three
lines would be inconsistent with the conclusion from SUMER
observations that the observed slow magnetoacoustic waves are
damped by thermal conduction \citep{Ofman2002}. Because of the
strong temperature dependence of the thermal conductivity, we
would expect the damping rate to increase with increasing
temperature \citep{Porter1994}.

Because the corona is threaded with a magnetic field, a diverse
array of oscillation modes is possible
\citep[e.g.,][]{Roberts2000,Roberts2003}. The sensitivity and
temperature coverage of EIS, especially when combined with the
high time cadence imaging that XRT can provide, opens up the
possibility of detecting new wave modes and new excitation
mechanisms. Similarly, this increased sensitivity may make it
possible to detect additional dissipation mechanisms. For
example, recent work by \citet{Ofman2007} and \citet{Selwa2007}
show that wave leakage may be important in the complex magnetic
geometries of active regions.

While EIS is capable of detecting Doppler shifts of of less than
0.5 km s$^{-1}$, the result is dependent on an accurate removal
of the orbital Doppler shift fluctuation, which has an amplitude
of about 20 km s$^{-1}$. Since the observation took place over
more than one orbit of the spacecraft, the orbital variation can
be accurately modelled and we believe it has been correctly
removed. In addition, the \textit{Hinode} pointing exhibits a
jitter in both the $x$ and $y$ directions that can have a range
of up to 4\arcsec\ over a 120~min observation. We have looked at
the jitter over the period of the sit-and-stare observation and
see no evidence for any correlation between the jitter and the
detected oscillations. Note, however, that this jitter means that
the EIS 1\arcsec\ slit is averaging over more than 1\arcsec\ in
both $x$ and $y$ during the observation. The fact that we see the
oscillation taking place over at least seven pixels in the
$y$-direction suggests that the oscillating structures are
sizable, and the jitter should not cause a significant problem.

To further verify this observation, we have performed the same
analysis on the oscillatory Doppler-shift signal marked in the
lower portions of Figure~\ref{fig:eis-sns-v}. The oscillations
observed over that position range (nine pixels), have similar
amplitudes, a shorter period (about 25~min), and similar damping
times. There is, however, less evidence for a temperatures
dependence to the ratio of the damping time to the period. We
believe that this second event in this EIS sit-and-stare data set
confirms the presence of these oscillations at active region
temperatures, but shows the need for additional studies on the
damping mechanism.

\acknowledgments \textit{Hinode} is a Japanese mission developed,
launched, and operated by ISAS/JAXA in partnership with NAOJ,
NASA, and STFC (UK). Additional operational support is provided
by ESA and NSC (Norway). JTM and HPW acknowledge support from the
NASA \textit{Hinode} program. We thank the anonymous referee for
suggestions that resulted in a much improved contribution.

\bibliographystyle{apj}
\bibliography{allrefs}

\begin{thebibliography}{16}
\expandafter\ifx\csname natexlab\endcsname\relax\def\natexlab#1{#1}\fi

\bibitem[{{Aschwanden} {et~al.}(1999){Aschwanden}, {Fletcher}, {Schrijver}, \&
  {Alexander}}]{Aschwanden1999}
{Aschwanden}, M.~J., {Fletcher}, L., {Schrijver}, C.~J., \& {Alexander}, D.
  1999, \apj, 520, 880

\bibitem[{{Culhane} {et~al.}(2007)}]{Culhane2007}
{Culhane}, J.~L., {et~al.} 2007, \solphys, 243, 19

\bibitem[{{Golub} {et~al.}(2007)}]{Golub2007}
{Golub}, L., {et~al.} 2007, \solphys, 243, 63

\bibitem[{{Kliem} {et~al.}(2002){Kliem}, {Dammasch}, {Curdt}, \&
  {Wilhelm}}]{Kliem2002}
{Kliem}, B., {Dammasch}, I.~E., {Curdt}, W., \& {Wilhelm}, K. 2002, \apjl, 568,
  L61

\bibitem[{{Kosugi} {et~al.}(2007)}]{Kosugi2007}
{Kosugi}, T., {et~al.} 2007, \solphys, 243, 3

\bibitem[{{Mariska}(2006)}]{Mariska2006}
{Mariska}, J.~T. 2006, \apj, 639, 484

\bibitem[{{Ofman}(2007)}]{Ofman2007}
{Ofman}, L. 2007, \apj, 655, 1134

\bibitem[{{Ofman} \& {Wang}(2002)}]{Ofman2002}
{Ofman}, L., \& {Wang}, T. 2002, \apjl, 580, L85

\bibitem[{{Porter} {et~al.}(1994){Porter}, {Klimchuk}, \&
  {Sturrock}}]{Porter1994}
{Porter}, L.~J., {Klimchuk}, J.~A., \& {Sturrock}, P.~A. 1994, \apj, 435, 482

\bibitem[{{Roberts}(2000)}]{Roberts2000}
{Roberts}, B. 2000, \solphys, 193, 139

\bibitem[{{Roberts} \& {Nakariakov}(2003)}]{Roberts2003}
{Roberts}, B., \& {Nakariakov}, V.~M. 2003, in NATO Science Series: II:
  Mathematics, Physics and Chemistry, Vol. 124, Turbulence, Waves and
  Instabilities in the Solar Plasma, ed. R.~Erdelyi, K.~Petrovay, B.~Roberts,
  \& M.~J. Aschwanden (Kluwer Academic Publishers, Dordrecht), 167--192

\bibitem[{{Sakurai} {et~al.}(2002){Sakurai}, {Ichimoto}, {Raju}, \&
  {Singh}}]{Sakurai2002}
{Sakurai}, T., {Ichimoto}, K., {Raju}, K.~P., \& {Singh}, J. 2002, \solphys,
  209, 265

\bibitem[{{Schrijver} {et~al.}(2002){Schrijver}, {Aschwanden}, \&
  {Title}}]{Schrijver2002}
{Schrijver}, C.~J., {Aschwanden}, M.~J., \& {Title}, A.~M. 2002, \solphys, 206,
  69

\bibitem[{{Selwa} {et~al.}(2007){Selwa}, {Ofman}, \& {Murawski}}]{Selwa2007}
{Selwa}, M., {Ofman}, L., \& {Murawski}, K. 2007, \apjl, 668, L83

\bibitem[{{Wang} {et~al.}(2002){Wang}, {Solanki}, {Curdt}, {Innes}, \&
  {Dammasch}}]{Wang2002}
{Wang}, T., {Solanki}, S.~K., {Curdt}, W., {Innes}, D.~E., \& {Dammasch}, I.~E.
  2002, \apjl, 574, L101

\bibitem[{{Wang} {et~al.}(2003){Wang}, {Solanki}, {Curdt}, {Innes}, {Dammasch},
  \& {Kliem}}]{Wang2003}
{Wang}, T.~J., {Solanki}, S.~K., {Curdt}, W., {Innes}, D.~E., {Dammasch},
  I.~E., \& {Kliem}, B. 2003, \aap, 406, 1105

\end{thebibliography}

\end{document}